# Valley-Selective Response of Nanoantennas Coupled to 2D Transition-Metal Dichalcogenides

Alex Krasnok[1], and Andrea Alù[2,1*]

[1]Department of Electrical and Computer Engineering, The University of Texas at Austin, Austin, Texas 78712, USA

[2]Advanced Science Research Center, City University of New York, New York, NY 10031, USA

E-mail: aalu@gc.cuny.edu

## Abstract

Monolayer (1L) transition metal dichalcogenides (TMDCs) are attractive materials for several optoelectronic applications because of their strong excitonic resonances and valley-selective response. Valley excitons in 1L-TMDCs are formed at opposite points of the Brillouin zone boundary, giving rise to a valley degree of freedom that can be treated as a pseudospin and may be used as a platform for information transport and processing. However, short valley depolarization times and relatively short exciton lifetimes at room temperature prevent using valley pseudospin in on-chip integrated valley devices. Recently it has been demonstrated how coupling these materials to optical nanoantennas and metasurfaces can overcome this obstacle. Here, we review the state-of-the-art advances in valley-selective directional emission and exciton sorting in 1L-TMDC mediated by nanostructures and nanoantennas. We briefly discuss the optical properties of 1L-TMDCs paying special attention to their photoluminescence/absorption spectra, dynamics of valley depolarization and valley Hall effect. Then, we review recent works on nanostructures for valley-selective directional emission from 1L-TMDCs.

## 1. Introduction

New directions in information processing and harvesting is of crucial importance in today's technology, giving rise to new fields of research, such as spintronics and valleytronics, in which spin or pseudospin are used as alternative information carriers. Currently, valleytronics is of great interest due to the discovery of atomically thin single layered (1L) transition metal dichalcogenides (TMDCs)[1–10]. Monolayer TMDCs are formed (Figure 1a) by a hexagonal network of transition metal atoms (M: Mo, W) hosted between two hexagonal lattices of chalcogenide atoms (X: S, Se). Electronically, 1L-TMDCs behave as two-dimensional semiconductors, with bandgaps lying in the visible and near-IR range. In the monolayer limit, the bandgaps of these materials are direct, enabling enhanced interactions of dipole transitions with light. An essential property of 1L-TMDCs is the broken inversion symmetry of their hexagonal 2D crystals, which, in combination with time-reversal symmetry, leads to opposite spins at the +K and -K valleys, effectively locking the spin and valley degrees of freedom (pseudospin or Berry curvature)[11–13]. Optical manipulation of the valley degree of freedom can be realized via exciton resonances based on the valley contrasting optical selection rules (for instance with $\sigma^-$ and $\sigma^+$ light excitation). The unique opportunity of addressing valley index make it possible to explore this binary quantum degree of freedom as an alternative information carrier [7,14,15], which may complement both classical and quantum computing schemes based on charge and spin.

However, short valley depolarization times and relatively short exciton lifetimes at room temperature prevent using valley pseudospin in on-chip integrated valley devices. Recently it has been demonstrated how coupling these materials to optical nanoantennas and metasurfaces can overcome this obstacle. It has been demonstrated that resonant optical nanostructures and nanoantennas can effectively enhance and direct the emission from opposite valleys in 1L-TMDCs into different directions [16–19]. The proposed nanostructures for valley-selective directional emission can be divided into two groups: nanostructures for *spatial separation* of valley degrees of freedom using surface waves and for *separation in K-space*, when photons with opposite helicity are emitted to different directions. Spatial separation of valley degrees of freedom may be accompanied with real *dividing of valley excitons in space* [19] or transformation of valley exciton pseudospin to photonic degrees of freedom (e.g., *transverse optical spin angular momentum*) via *optical spin-orbit coupling* [20] (or both these processes).

This paper is devoted to reviewing state-of-the-art advances in 1L-TMDC valley degree of freedom separation (valley-selective directional emission and exciton sorting) mediated by nanostructures and nanoantennas. The paper is organized as follows. In Section 2 we briefly discuss the optical properties of 2D TMDC materials paying special attention to their photoluminescence/absorption spectra, dynamics of valley depolarization and valley Hall effect. In Section 3 we summarize recent work devoted to nanostructures for 1L-TMDC valley separation published. Special attention is devoted to our and our collaborators' recent results in this area.

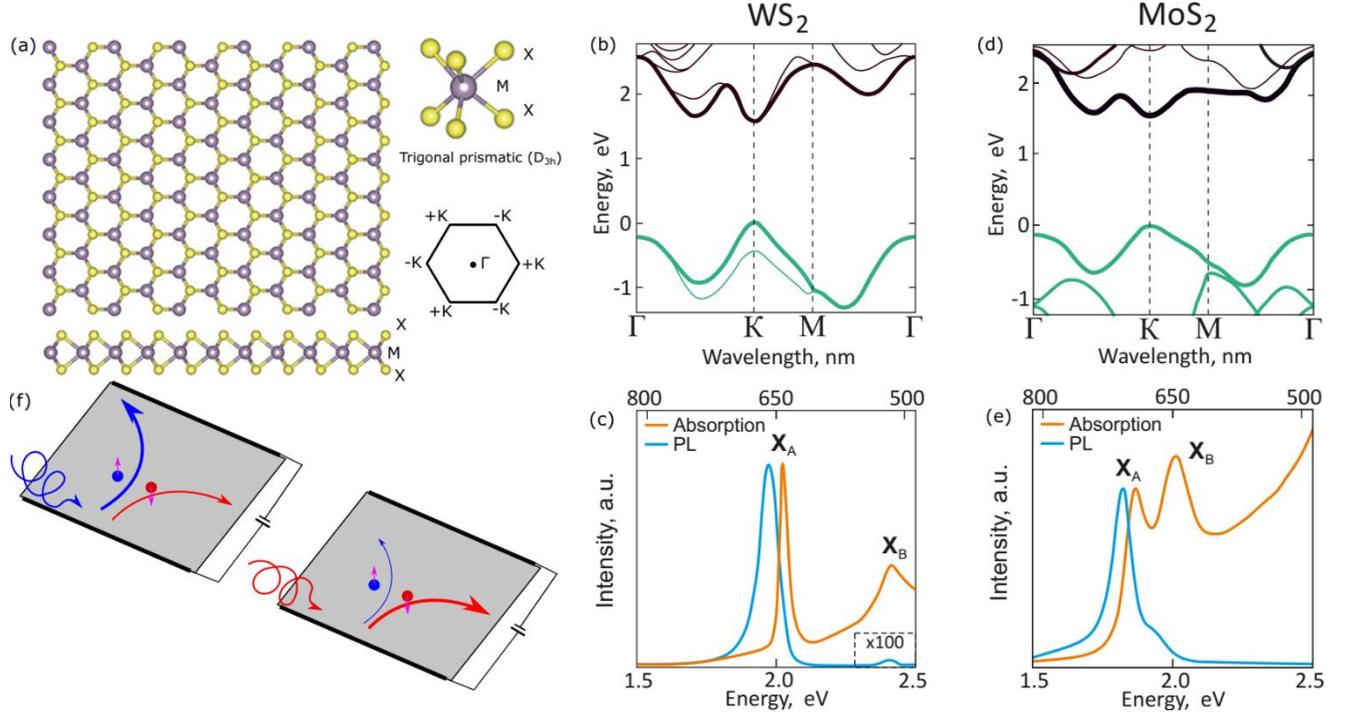

**Figure 1**. (a) Monolayer transition metal dichalcogenide crystal structure. Transition metal atoms (M) appear in purple, chalcogen atom (X) in yellow. The structure resembles the one of graphene, but with broken inversion symmetry. A side view shows the 3D structure. The hexagonal Brillouin zone is shown with the Γ point and the two inequivalent +K and −K points. (b) Energy structure of 1L-WS$_2$, blue and black curves demonstrate the valence and conduction bands. (c) Absorption spectra (orange curve) and photoluminescence (blue curve) of 1L-WS$_2$ at room temperature. (d), (e) The same for MoS$_2$. (f) Valley Hall effect. Blue (red) denotes the electron in valley +K (−K). The purple arrows indicate the pseudo-vector quantities (Berry curvature or orbital magnetic moment) of the electron (hole); blue and red spirals indicate $\sigma^-$ and $\sigma^+$ light excitation.

## 2. Optical properties of 1L-TMDCs

Single layered TMDCs (like MoS$_2$, MoSe$_2$, WS$_2$, WSe$_2$) behave as two-dimensional semiconductors, with their bandgaps lying in the visible and near-IR ranges [2]. In the monolayer limit, TMDCs are particularly interesting because their bandgaps become direct, thus enabling enhanced interactions of dipole transitions with light [21]. The transition to a direct bandgap semiconductor makes it easier for photons with an energy equal to the bandgap to be absorbed or emitted, without requiring additional phonons to be absorbed. As a result, the resonant excitonic absorption of a single TMD monolayer can be as large as 5–20%, which is one order of magnitude larger than, for instance, the one of quantum well systems. For example, Figures 1b-e demonstrate energy structures of 1L-WS2 (b) and 1L-MoS2 (d) as well as their absorption spectra and photoluminescence (c, e) at room temperature. The PL spectrum consists of two resonances corresponding to $X_A$ and $X_B$ excitons, but the PL emission from $X_B$ exciton is very weak. We note that the optical properties of 1L-TMDCs (PL spectra, Raman spectra, transmission/absorption spectra) may slightly depend on several factors, including fabrication techniques (mechanical exfoliation or chemical vapor deposition), type of substrate, defects and others.

Reduced dielectric screening and strong Coulomb interactions between charged particles (electrons and holes) result in strong excitonic resonances in the visible and near-IR range with large binding energies. Strong Coulomb interactions in these materials lead to the formation of strongly bound excitons (binding energies of 0.2 to 0.8 eV)[22–24], charged excitons (trions)[25,26], and excitonic molecules (biexcitons) [19–23]. Localized and dark excitonic states in 1L-TMDCs have also been demonstrated [19,24,25]. Due to the monolayer nature of 1L-TMDCs, their high oscillator strength and the potential for tuning, these materials have become a unique class of 2D materials for optoelectronic applications, such as photodetection and light harvesting[34–37], phototransistors and modulation[38,39] and light-emitting diodes[40–42]. Operation of 1L-TMDCs in single photon emission regime has also been reported[43,44]. The enhancement of the exciton transition dipole in 1L-TMDCs can be understood assuming that it is proportional to the electron-hole overlap in real space. Recently, it has been demonstrated that the bright exciton dipole moment can be further enhanced via dielectric screening effect in high-dielectric constant solvents (like ethanol and water), which gives rise in asymmetric Fano resonance and strong coupling effects[45]. The quantum yield of emission of these materials depend on many factors, including fabrication techniques (mechanical exfoliation or chemical vapor deposition), type of substrate, defects and so on. Typically, as-prepared 1L-TMDCs have a relatively low quantum yield of ~0.1-10% [46]. These low values can be significantly improved (up to ~95%) by chemical treatment by organic superacids [47,48] or by coupling them to specifically tailored resonant *optical nanocavities and metasurfaces* [33,49–63].

A variety of new optical effects stemming from the interaction of 1L-TMDCs with plasmonic (i.e., made of noble metals) and high-index dielectric (Si, Ge, GaP) nanocavities has been demonstrated. Examples include the observation of strong plasmon-exciton coupling [56,63–65], pronounced Fano resonances [51] and plasmon-induced resonance energy transfer [66], which are very attractive for various quantum optics and nanophotonic applications. These effects benefit from localization of light within plasmonic and dielectric resonators, enabled by their small mode volumes and the strong dipole moment of excitons in 1L-TMDCs.

Direct band gap transitions in 1L-TMDCs occur at the energy-degenerate K (K') points at the edges of the 2D hexagonal Brillouin zone (schematically shown in Figure 1a). Due to inversion symmetry breaking and strong spin-orbital coupling in 1L-TMDCs, the electronic states of the two valleys have different chirality, which leads to valley-selective circular dichroism [13,67–71]. This effect is key for valleytronics applications, which focus on the manipulation of valley pseudo-spins to encode signals and information [7,20,71,72]. The operation principles of valleytronics are mainly based on the *valley Hall effect* [68], which refers to the opposite Hall currents for carriers located in opposite valleys, Figure 1f. Excitation of 1L-TMDCs by $\sigma^+$ and $\sigma^-$ light causes nonequilibrium population of +K or -K valley. These carriers have opposite Berry curvature, which acts like effective magnetic field in the momentum space. As a result, these two sort of valley particles (+K or -K) propagate to different directions. If the particles are electrically charged (electrons, holes, trions) this effect gives rise to the Hall voltage. If the particles are neutral, the Hall effect can be detected by measuring the circularly polarized components of the emitted photoluminescence (PL) intensity. In this case the effect can be characterized by *degree of valley polarization*

$$\rho(\omega) = \frac{I_+(\omega) - I_-(\omega)}{I_+(\omega) + I_-(\omega)}, \qquad (1)$$

where $I_+(\omega)$ and $I_-(\omega)$ refers to the PL intensity with $\sigma^+$ and $\sigma^-$ polarization. The Hall effect can be also detected in reflectance experiments via so-called *Kerr rotation spectroscopy*, when a system illuminated by a linearly polarized light causes rotation of polarization in scattered signal [73–75]. By spatially resolving the Kerr signal, the valley polarization can be mapped out in a sample, and accumulation of +K and −K valley polarizations can be observed.

Free carrier valley Hall effect in monolayer MoS2 was first demonstrated using electric readout methods [68]. There, the Berry curvature of the energy band acts as a momentum-dependent magnetic field, leading to transverse motions of valley polarized carriers in the presence of an in-plane electric field. More recently, exciton valley Hall effect was reported for a monolayer $MoS_2$ where the exciton motion is driven by a temperature gradient[71]. Moreover, other factors such as uncontrolled strain can also lead to the separation of valley-polarized excitons.

For a full description and understanding of the 1L-TMDCs behavior, it is important to know the exciton and coherence lifetimes [76]. The exciton lifetime describes the average time during which an exciton exists, and it usually defines the excitonic spectral linewidth. This time also limits the propagation distance of excitons. The coherence time defines the time during which an exciton remembers the state of the excitation field (for example, polarization). If the system is excited by $\sigma^+$ or $\sigma^-$ light, the coherence lifetime says how long the +K or −K valley are inequivalently polarized (nonzero valley polarization). In 1L-TMDCs the coherence time is usually less than the exciton lifetime (at room temperature) and it is key for valleytronics and quantum optics applications [77]. The dynamics of photoexcited electron−hole pairs in 1L-WSe$_2$ at room temperature and ambient conditions has been directly traced in Ref. [78]. It has been shown that after highly nonresonant interband excitation by the femtosecond laser pulse, the concentration of free carriers increases, reaching its maximum after 0.5 fs. This excitation is followed by a rapid carrier relaxation towards the respective band minima. More than half of the carriers are bound into excitons already 0.4 ps after the excitation. The ratio between excitons and unbound electron-hole pairs increases up to 0.5 ps. Interestingly, the exciton concentration grows even after 0.5-0.6 ps, when the free carrier concentration starts to decay, and continues up to 1 ps. Then, both concentrations decay on a time scale of a few picoseconds, while a significant fraction of free carriers is still observed after 5 ps. We note that the results for cryogenic temperatures are similar because of strong bounding energies (~0.5 eV) of excitons, which are well above the thermal energy (~50 meV).

The valley polarization degree dynamics of 1L-WSe$_2$ has been experimentally studied in Ref. [79] by pump-probe Kerr rotation technique. It has been demonstrated that the exciton valley depolarization time decreases significantly from ~6 ps up to ~1.5 ps when the lattice temperature increases from 4 K up to 125 K, which is in a good agreement with theoretical estimations[80]. In Ref. [81] these results have been obtained from *ab initio* calculations, which took into account all possible mechanisms of valley depolarization. Similar results have been demonstrated in [32]. At room temperatures the valley depolarization times are well less than 1 ps, which is much less than exciton lifetimes. Thus, all previously reported carrier and exciton valley Hall effects relying on nonzero valley polarization were observed at low temperatures. At room temperature the valley polarization lifetime is too short to be utilized in real applications. However, since the binding energy of excitons is very large in 1L-TMDCs, it is therefore both feasible and highly desirable to design a photonic device to manipulate valley excitons at room temperature even without valley polarization. These structures will be reviewed in the next section.

## 3. Nanostructures for Valley-Selective Directional Emission

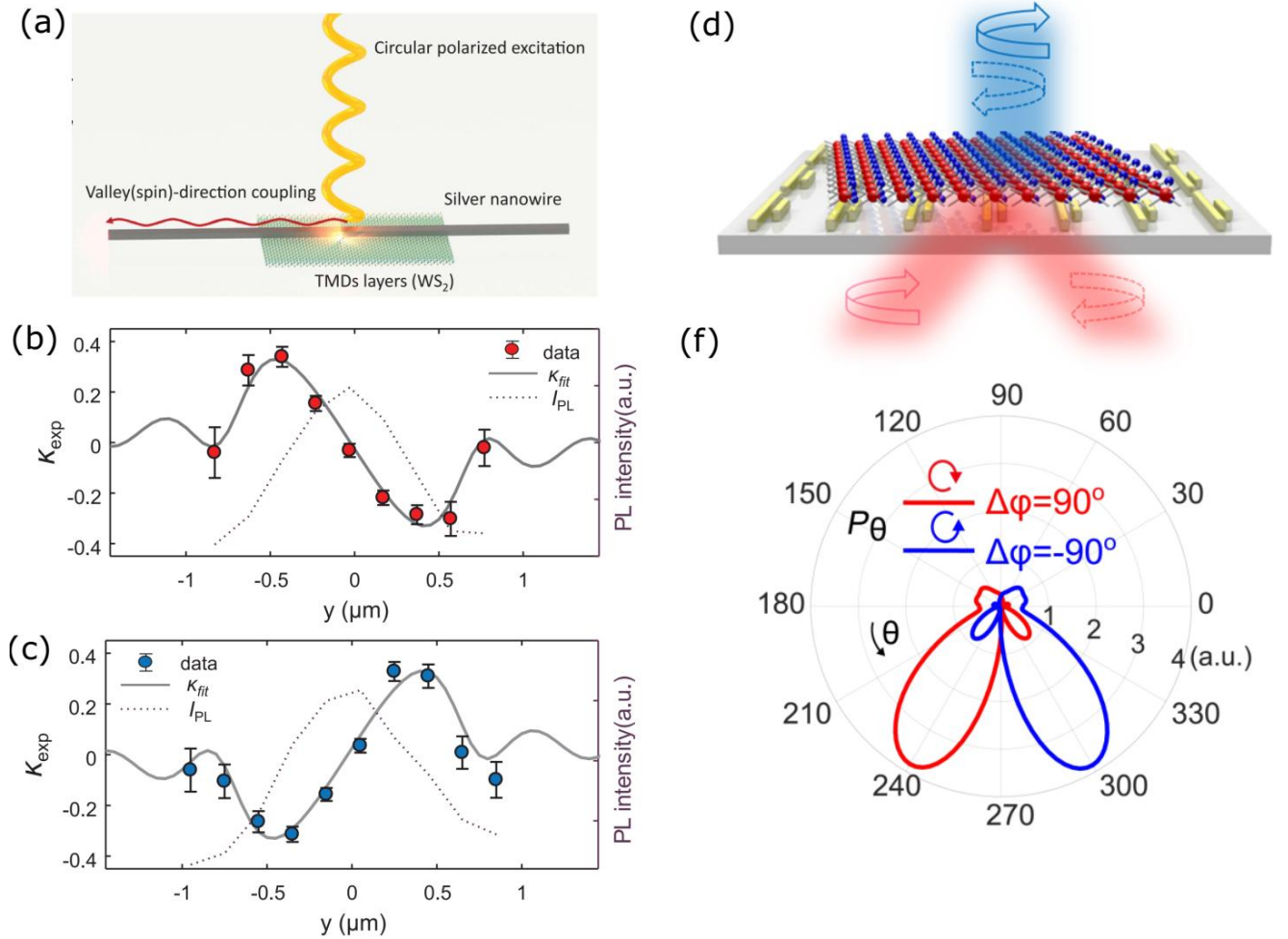

**Figure 2**. A conceptual illustration of directional emission of a valley-polarized exciton in WS$_2$. The valley pseudospin and photon path are coupled by means of spin-orbit coupling of light [20]. (b), (c) Measured directional coupling efficiency, $(I_L - I_R)/(I_L + I_R)$, of the guided emission as a function of the position of the excitation 594-nm laser in the transverse direction (y) with the (b) left- and (c) right-handed circular polarization. Gray lines represent fitting results obtained by using the calculated directional coupling efficiency. Purple dotted lines correspond to the total PL intensity measured from the ends of the nanowire as a function of the excitation position. (d) Array of double-bar

plasmonic antennas for valley-selective directional emission from a 1L-TMDC. (f) Polar plots of the radiated power in the far field; $\Delta\varphi = \pm 90°$ denote right and left circularly polarized excitons [16].

## 3.1. Single nanoantennas

As mentioned in the introduction, the proposed nanostructures for valley-selective response can perform spatial separation of valley degree of freedom with surface waves or separation in K-space when photons with opposite helicity are emitted to different directions. Moreover, spatial separation of the valley degree of freedom may be accompanied by real dividing of valley excitons [19] or transformation of valley exciton pseudospin to photonic degrees of freedom via optical spin-orbit coupling [17,20,82–85] (or both these process). Below we desribe all these opportunities with various examples from the recent literature.

Photons as information carriers are more robust and controllable than excitons, especially at room temperature. Thus, nanostructures supporting coupling of valley pseudospin to photonic degrees of freedom (e.g., transverse optical spin angular momentum, t-OSAM) are desirable for on-chip integrated valley devices. The optical spin-orbit coupling effect provides a robust one-to-one transformation of valley degree of freedom to t-OSAM because of spin-momentum locking. One realization of this idea has been implemented in Ref. [20], Figure 2a. The structure consists of a single silver nanowire placed on top of a few-layer TMD material (WS2). The silver nanowire plasmonic supports guided modes possessing t-OSAM in outplane direction. The spin-momentum locking of these plasmonic modes allows to transformation of valley polarization into directional emission: the system being excited by circularly polarized light in the middle point of the nanowire launches the plasmonic modes carrying t-OSAM to the left or right direction. Note that because of the structure symmetry, the asymmetric operation can be achieved under asymmetric laser spot excitation in the transverse direction (y). This effect can be described by coupling efficiency $(I_L - I_R)/(I_L + I_R)$, where $I_L$ and $I_R$ are light intensities directed to the left and right side, respectively. Thus, when this value is zero, the structure demonstrated equivalent coupling to the both sides. The measurement results of the coupling efficiency for 594-nm $\sigma^+$ and $\sigma^-$ light excitation as a function of the laser spot coordinate in the transverse direction are shown in Figure 2b and c. The purple dotted lines show the total PL intensity measured from the ends of the nanowire as a function of the excitation position.

Plasmonic (Au) nanoantennas for valley dependent far-field emission (separation in K-space) when photons with opposite helicity are emitted to different directions have been theoretically proposed in [16], Figure 2d. The antenna consists of two closely arranged plasmonic bars different in their lengths. Short bar possesses electric dipole resonance, while the longer one has quadrupole resonance at the same operation wavelength. The entire nanoantenna being excited by a circularly polarized dipole (modeled as two orthogonal short dipoles with the relative phase of $\Delta\varphi = \pm 90°$) emits in different directions. The asymmetry originates from constructive or destructive interference of fields from dipole and quadrupole resonances. Thus, when the coupled 1L-TMDC–nanoantenna system is excited by light of different circular polarizations, the emission from the two valleys (+K and -K) will be emitted into opposite directions (Figure 2f).

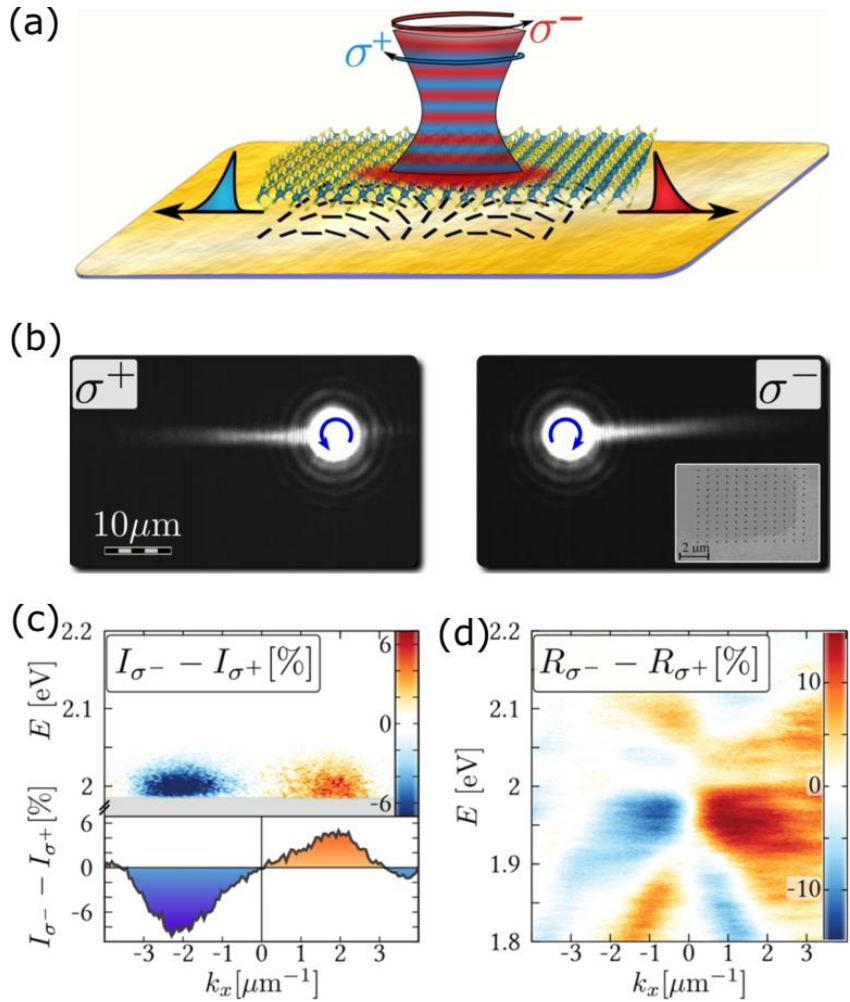

**Figure 3**. (a) Chiral coupling of valley excitons in a 1L-TMDC with spin-momentum locked surface plasmons on a surface with array with $\varphi$-rotated rectangular apertures [17]. (b) Real-space leakage radiation microscope images of the surface plasmons launched by $\sigma^-$ and $\sigma^+$ excitations. (c)

Differential PL dispersion spectrum for left and right circularly polarized excitations. The shaded regions in all panels are removed by the laser line filter, and the cross-cuts are taken at 2 eV. (d) Differential angle-resolved reflection spectrum for left and right circularly polarized light.

## 3.2. Metasurfaces

Metasurfaces for valley-selective response of 2D transition-metal dichalcogenides at room temperature have been proposed in Refs. [17,19]. The metasurface proposed in Ref. [17] consists of an array of rectangular nanoapertures in a 200-nm film of gold arranged with a grating period $\Lambda$ rotated stepwise by an angle $\pi/6$ (Figure 3a). The resulting metasurface is covered by mechanically exfoliated monolayer of WS2. The metasurface chirality in plane allows valley-selective spin-momentum locked propagation of the PL signal (Figure 3b). Moreover, it has been shown that the proposed resonant metasurface shows a coupling regime, which is very close to the strong couplings [86,87] with Rabi splitting energy of 40 meV at the crossing point (A-exciton emission energy, 2.01 eV) between the uncoupled exciton and the surface mode of the metasurface. As a result, the entire structure forms hybrid exciton-photon quasiparticles (referred to as chiralitons). In this case the spatial separation of valley degree of freedom is accompanied by real dividing of valley excitons in space in for of polaritons. The difference between PL dispersions obtained with left and right circularly polarized excitations is displayed in Figure 3c, showing net flows of chiralitons with spin-determined momenta. The hybrid structure shows also resonant reflection with pronounced asymmetry in differential white light reflectivity maps (Figure 3d).

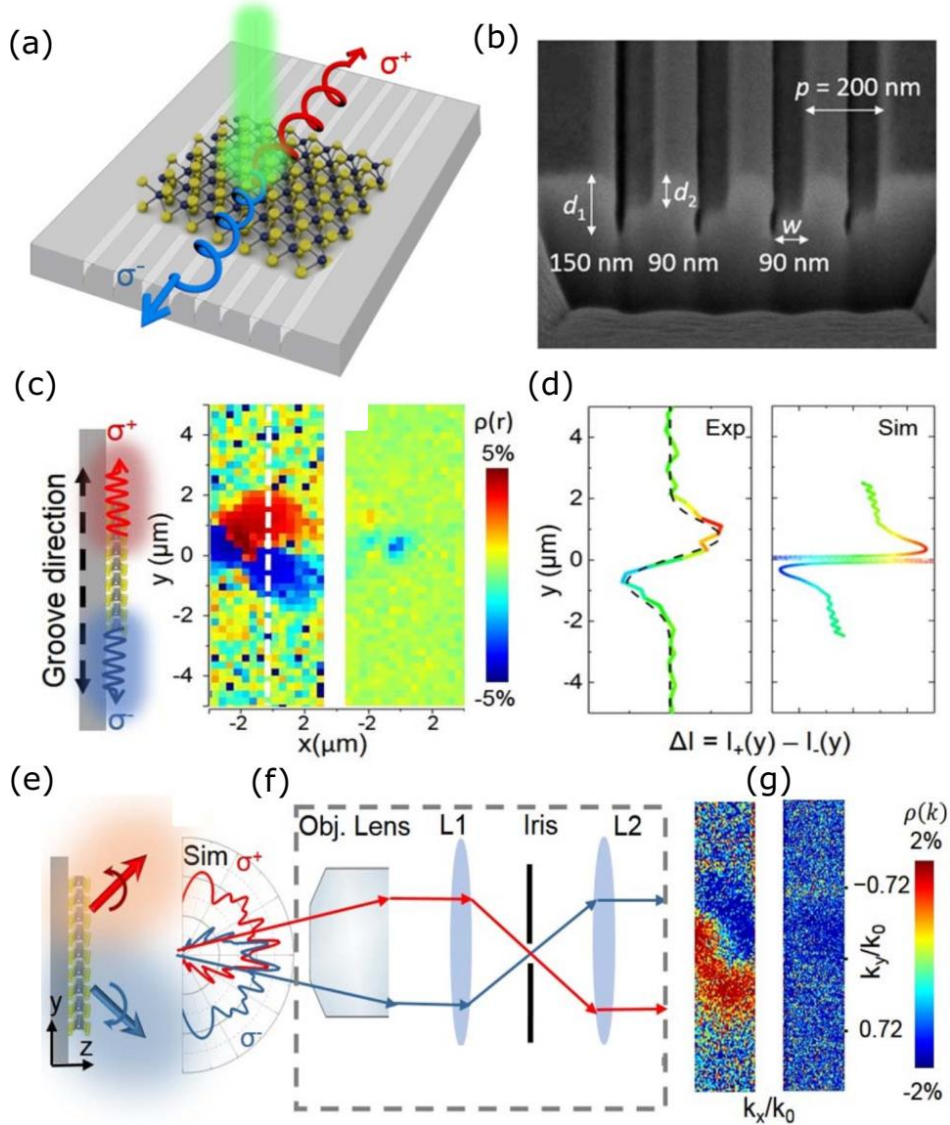

**Figure 4.** Routing valley excitons in a 1L-MoS2 with a metasurface [19]. (a) Illustration of valley excitons in a monolayer TMD controlled by a metasurface consisting of asymmetric grooves. Not only valley polarized excitons are spatially separated, photons with opposite helicity are also emitted to different directions, serving as a valley-photon interface mediated by excitons. (b) SEM image of the cross section of the asymmetric grooves. (c) Illustration of valley exciton separation caused by the metasurface. Real space color plot of valley polarization obtained from the (left) MoS2- metasurface and (right) MoS2-flat silver film under 532-nm laser excitation. (d) A line profile $\Delta I(y) = I_+(y) - I_-(y)$ along the groove. Black dash curve is a fitting by subtracting two Gaussian curves. (right) A line profile from simulation. (e) Illustration of the helicity-dependent directional emission of the valley excitons. (right) Numerically simulated far-field emission pattern from valley excitons with opposite helicity. (f) Set-up for the k-space mapping of PL. (g) Experimental $\rho(k_\parallel)$

distribution in photon momentum space obtained from (left) MoS2-metasurface (right) MoS2-flat silver film.

Another realization of a metasurface for valley exciton sorting and routing in a 1L-MoS2 has been proposed in Ref. [19]. The metasurface consists of asymmetrically shaped grooves on a silver (Ag) film arranged in a subwavelength period, Figure 4a. The SEM image of the cross section of the realized asymmetric grooves is shown in Figure 4b. The spatial separation of valley excitons of opposite chirality is enabled by coupling to surface plasmon polaritons (SPPs) propagating along the asymmetrically shaped grooves. As already mentioned, the +K and -K valley excitons in a 1L-TMDCs can be modeled as in-plane circularly polarized dipoles oscillating with opposite helicity. However, their in-plane oriented dipoles cannot asymmetrically excite conventional SPPs, since they do not engage the required out-of-plane chirality. Moreover, the efficiency of in-plane dipoles coupling to TM polarized SPPs is very small. These challenges require metasurfaces with asymmetric grooves, where the effective coupling between 1L-TMDC excitons and SPP waves is possible on the side walls. This crucial concept extends the photonic spin-Hall effect to planar metasurfaces with only in-plane electric field, and it enables chirality dependent coupling between TMDs and metasurfaces. Figure 4c shows the experimentally observed degree of valley polarization (left) pattern over the metasurface. These results demonstrate the asymmetry of measured PL signal from 1L-MoS2 on metasurfaces away from the laser excitation spot. The right plot shows the results from 1L-MoS2 on Ag homogeneous substrate, where the vanishing degree of valley polarization is observed. The valley exciton dividing of ~2 $\mu$m is demonstrated. Figure 4d demonstrates a line profile $\Delta I(y) = I_+(y) - I_-(y)$ measured (left) and simulated (right) along the groove. Black dash curve shows the fitting by subtracting two Gaussian curves.

It has been also demonstrated that the metasurface emits photons with opposite helicity to different directions, dividing the valley degree of freedom in far-field, Figure 4e. Numerical results (Figure 4e, right panel) show that the $\sigma^+$ and $\sigma^-$ polarized photons preferentially emit toward the upper and lower hemisphere. The experimental results of measurement of the valley degree of freedom in the K-space, obtained with a setup sketched in Figure 4f, are summarized in Figure 4g. These results clearly show the ability of the metasurface to divide the +K and -K valley excitons emission into different radiation directions.

Concluding this section, we note that the proposed solutions to separate valley index is rather general, and it can be applied to a wide range of layered materials. Moreover, no exciton valley polarization is required, as is the case for MoS2 at room temperature, which allows utilizing this approach in valleytronics applications at room temperature.

## Conclusions

In this paper, we have reviewed state-of-the-art advances in 1L-TMDC valley degree of freedom separation accompanied by valley-selective directional emission or real exciton dividing mediated by resonant plasmonic nanostructures and nanoantennas. These works are aimed to overcome a significant obstacle in the field of valleytronics related to short valley depolarization times and relatively short exciton lifetimes at room temperature, which prevent using valley pseudospin in on-chip integrated valley devices. The observed nanostructures for valley-selective directional emission allow either spatial separation of valley degree of freedom with surface waves or separation in K-space when photons with opposite helicity are emitted to different directions. The existing works show that spatial separation of valley degree of freedom may be accompanied by real dividing of valley excitons in space or transformation of valley exciton pseudospin to photonic degree of freedom (e.g. transverse optical spin angular momentum) via optical spin-orbit coupling (or both these process). This analysis is supported by a brief review of optical properties of 2D TMDC materials with attention on their photoluminescence/absorption spectra, dynamics of valley depolarization and valley Hall effect.

## Acknowledgements

This work was supported by the Air Force Office of Scientific Research and the Welch Foundation with grant No. F-1802.

## References

1. Butler, S. Z.; Hollen, S. M.; Cao, L.; Cui, Y.; Gupta, J. A.; Gutiérrez, H. R.; Heinz, T. F.; Hong, S. S.; Huang, J.; Ismach, A. F.; Johnston-Halperin, E.; Kuno, M.; Plashnitsa, V. V.; Robinson, R. D.; Ruoff, R. S.; Salahuddin, S.; Shan, J.; Shi, L.; Spencer, M. G.; Terrones, M.; Windl, W.; Goldberger, J. E. Progress, challenges, and opportunities in two-dimensional materials beyond graphene. *ACS Nano* **2013**, *7*, 2898–2926, doi:10.1021/nn400280c.

2. Xia, F.; Wang, H.; Xiao, D.; Dubey, M.; Ramasubramaniam, A. Two-dimensional material

nanophotonics. *Nat. Photonics* **2014**, *8*, 899–907, doi:10.1038/nphoton.2014.271.

3.  Mak, K. F.; Shan, J. Photonics and optoelectronics of 2D semiconductor transition metal dichalcogenides. *Nat. Photonics* **2016**, *10*, 216–226, doi:10.1038/nphoton.2015.282.

4.  Bhimanapati, G. R.; Lin, Z.; Meunier, V.; Jung, Y.; Cha, J.; Das, S.; Xiao, D.; Son, Y.; Strano, M. S.; Cooper, V. R.; Liang, L.; Louie, S. G.; Ringe, E.; Zhou, W.; Kim, S. S.; Naik, R. R.; Sumpter, B. G.; Terrones, H.; Xia, F.; Wang, Y.; Zhu, J.; Akinwande, D.; Alem, N.; Schuller, J. A.; Schaak, R. E.; Terrones, M.; Robinson, J. A. Recent Advances in Two-Dimensional Materials beyond Graphene. *ACS Nano* **2015**, *9*, 11509–11539, doi:10.1021/acsnano.5b05556.

5.  Das, S.; Robinson, J. A.; Dubey, M.; Terrones, H.; Terrones, M. Beyond Graphene: Progress in Novel Two-Dimensional Materials and van der Waals Solids. *Annu. Rev. Mater. Res.* **2015**, *45*, 1–27, doi:10.1146/annurev-matsci-070214-021034.

6.  Grosso, G. 2D materials: Valley polaritons. *Nat. Photonics* **2017**, *11*, 455–456, doi:10.1038/nphoton.2017.135.

7.  Schaibley, J. R.; Yu, H.; Clark, G.; Rivera, P.; Ross, J. S.; Seyler, K. L.; Yao, W.; Xu, X. Valleytronics in 2D materials. *Nat. Rev. Mater.* **2016**, *1*, 16055, doi:10.1038/natrevmats.2016.55.

8.  Manzeli, S.; Ovchinnikov, D.; Pasquier, D.; Yazyev, O. V.; Kis, A. 2D transition metal dichalcogenides. *Nat. Rev. Mater.* **2017**, *2*, 17033, doi:10.1038/natrevmats.2017.33.

9.  Koperski, M.; Molas, M. R.; Arora, A.; Nogajewski, K.; Slobodeniuk, A. O.; Faugeras, C.; Potemski, M. Optical properties of atomically thin transition metal dichalcogenides: observations and puzzles. *Nanophotonics* **2017**, *6*, 1–20, doi:10.1515/nanoph-2016-0165.

10. Krasnok, A.; Lepeshov, S.; Alú, A. Nanophotonics with 2D Transition Metal Dichalcogenides. *ArXiv* **2018**, *1801.00698*.

11. Mak, K. F.; He, K.; Shan, J.; Heinz, T. F. Control of valley polarization in monolayer MoS2 by optical helicity. *Nat. Nanotechnol.* **2012**, *7*, 494–498, doi:10.1038/nnano.2012.96.

12. Zeng, H.; Dai, J.; Yao, W.; Xiao, D.; Cui, X. Valley polarization in MoS2 monolayers by optical pumping. *Nat. Nanotechnol.* **2012**, *7*, 490–493, doi:10.1038/nnano.2012.95.

13. Cao, T.; Wang, G.; Han, W.; Ye, H.; Zhu, C.; Shi, J.; Niu, Q.; Tan, P.; Wang, E.; Liu, B.; Feng, J. Valley-selective circular dichroism of monolayer molybdenum disulphide. *Nat. Commun.* **2012**, *3*, 885–887, doi:10.1038/ncomms1882.

14. Xu, X.; Yao, W.; Xiao, D.; Heinz, T. F. Spin and pseudospins in layered transition metal dichalcogenides. *Nat. Phys.* **2014**, *10*, 343–350, doi:10.1038/nphys2942.

15. Xiao, D.; Liu, G. Bin; Feng, W.; Xu, X.; Yao, W. Coupled Spin and Valley Physics in Monolayers of MoS2 and Other Group-VI Dichalcogenides. *Phys. Rev. Lett.* **2012**, *108*, 196802, doi:10.1103/PhysRevLett.108.196802.

16. Chen, H.; Liu, M.; Xu, L.; Neshev, D. N. Valley-selective directional emission from a transition-metal dichalcogenide monolayer mediated by a plasmonic nanoantenna. *Beilstein J. Nanotechnol.* **2018**, *9*, 780–788, doi:10.3762/bjnano.9.71.

17. Chervy, T.; Azzini, S.; Lorchat, E.; Wang, S.; Gorodetski, Y.; Hutchison, J. A.; Berciaud, S.; Ebbesen, T. W.; Genet, C. Spin-momentum locked polariton transport in the chiral strong coupling regime. **2017**, 1–12.

18. Gong, S.-H.; Alpeggiani, F.; Sciacca, B.; Garnett, E. C.; Kuipers, L. Nanoscale chiral valley-photon interface through optical spin-orbit coupling. *Science (80-. ).* **2018**, *359*, 443–447, doi:10.1126/science.aan8010.

19. Liuyang Sun, Chun-Yuan Wang, Alexandr Krasnok, Junho Choi, Jinwei Shi, Juan Sebastian Gomez-Diaz, Andre' Zepeda, Shangjr Gwo, Chih-Kang Shih1, Andrea Alù, X. L. Routing Valley Excitons in a Monolayer MoS2 with a Metasurface. *Arxiv* **2018**, *1801.06543*.

20. Gong, S.; Alpeggiani, F.; Sciacca, B.; Garnett, E. C.; Kuipers, L. Nanoscale chiral valley-photon interface through optical spin-orbit coupling. *Science (80-. ).* **2018**, *359*, 443–447, doi:10.1126/science.aan8010.

21. Zhao, W.; Ghorannevis, Z.; Chu, L.; Toh, M.; Kloc, C.; Tan, P.-H.; Eda, G. Evolution of Electronic Structure in Atomically Thin Sheets of WS 2 and WSe 2. *ACS Nano* **2013**, *7*, 791–797, doi:10.1021/nn305275h.

22. Li, Y.; Chernikov, A.; Zhang, X.; Rigosi, A.; Hill, H. M.; van der Zande, A. M.; Chenet, D. A.;

Shih, E.-M.; Hone, J.; Heinz, T. F. Measurement of the Optical Dielectric Function of Monolayer Transition-Metal Dichalcogenides: MoS2, MoSe2, WS2, and WSe2. *Phys. Rev. B* **2014**, *90*, 205422, doi:10.1103/PhysRevB.90.205422.

23. Ye, Z.; Cao, T.; O'Brien, K.; Zhu, H.; Yin, X.; Wang, Y.; Louie, S. G.; Zhang, X.; O'Brien, K.; Zhu, H.; Yin, X.; Wang, Y.; Louie, S. G.; Zhang, X. Probing excitonic dark states in single-layer tungsten disulphide. *Nature* **2014**, *513*, 214–218, doi:10.1038/nature13734.

24. Ugeda, M. M.; Bradley, A. J.; Shi, S.-F.; da Jornada, F. H.; Zhang, Y.; Qiu, D. Y.; Ruan, W.; Mo, S.; Hussain, Z.; Shen, Z.; Wang, F.; Louie, S. G.; Crommie, M. F. Giant bandgap renormalization and excitonic effects in a monolayer transition metal dichalcogenide semiconductor. *Nat. Mater.* **2014**, *13*, 1091–1095, doi:10.1038/nmat4061.

25. Ross, J. S.; Wu, S.; Yu, H.; Ghimire, N. J.; Jones, A. M.; Aivazian, G.; Yan, J.; Mandrus, D. G.; Xiao, D.; Yao, W.; Xu, X. Electrical control of neutral and charged excitons in a monolayer semiconductor. *Nat. Commun.* **2013**, *4*, 1473–1476, doi:10.1038/ncomms2498.

26. Cuadra, J.; Baranov, D. G.; Wersäll, M.; Verre, R.; Antosiewicz, T. J.; Shegai, T. Observation of tunable charged exciton polaritons in hybrid monolayer WS2 – plasmonic nanoantenna system. *arXiv* **2017**, 1703.07873.

27. Shang, J.; Shen, X.; Cong, C.; Peimyoo, N.; Cao, B.; Eginligil, M.; Yu, T. Observation of excitonic fine structure in a 2D transition-metal dichalcogenide semiconductor. *ACS Nano* **2015**, *9*, 647–655, doi:10.1021/nn5059908.

28. Sie, E. J.; Lui, C. H.; Lee, Y. H.; Kong, J.; Gedik, N. Observation of Intervalley Biexcitonic Optical Stark Effect in Monolayer WS2. *Nano Lett.* **2016**, *16*, 7421–7426, doi:10.1021/acs.nanolett.6b02998.

29. You, Y.; Zhang, X. X.; Berkelbach, T. C.; Hybertsen, M. S.; Reichman, D. R.; Heinz, T. F. Observation of biexcitons in monolayer WSe 2. *Nat. Phys.* **2015**, *11*, 477–481, doi:10.1038/nphys3324.

30. Hao, K.; Specht, J. F.; Nagler, P.; Xu, L.; Tran, K.; Singh, A.; Dass, C. K.; Schüller, C.; Korn, T.; Richter, M.; Knorr, A.; Li, X.; Moody, G. Neutral and charged inter-valley biexcitons in monolayer MoSe2. *Nat. Commun.* **2017**, *8*, 15552, doi:10.1038/ncomms15552.

31. Kim, M. S.; Yun, S. J.; Lee, Y.; Seo, C.; Han, G. H.; Kim, K. K.; Lee, Y. H.; Kim, J. Biexciton Emission from Edges and Grain Boundaries of Triangular WS2 Monolayers. *ACS Nano* **2016**, *10*, 2399–2405, doi:10.1021/acsnano.5b07214.

32. Plechinger, G.; Korn, T.; Lupton, J. M. Valley-Polarized Exciton Dynamics in Exfoliated Monolayer WSe 2. *J. Phys. Chem. C* **2017**, *121*, 6409–6413, doi:10.1021/acs.jpcc.7b01468.

33. Park, K.-D.; Jiang, T.; Clark, G.; Xu, X.; Raschke, M. B. Radiative control of dark excitons at room temperature by nano-optical antenna-tip Purcell effect. *Nat. Nanotechnol.* **2017**, 1–7, doi:10.1038/s41565-017-0003-0.

34. Lopez-Sanchez, O.; Lembke, D.; Kayci, M.; Radenovic, A.; Kis, A. Ultrasensitive photodetectors based on monolayer MoS2. *Nat. Nanotechnol.* **2013**, *8*, 497–501, doi:10.1038/nnano.2013.100.

35. Britnell, L.; Ribeiro, R. M.; Eckmann, A.; Jalil, R.; Belle, B. D.; Mishchenko, A.; Kim, Y.-J.; Gorbachev, R. V.; Georgiou, T.; Morozov, S. V.; Grigorenko, A. N.; Geim, A. K.; Casiraghi, C.; Neto, A. H. C.; Novoselov, K. S. Strong Light-Matter Interactions in Heterostructures of Atomically Thin Films. *Science (80-. ).* **2013**, *340*, 1311–1314, doi:10.1126/science.1235547.

36. Perea-Lõpez, N.; Elías, A. L.; Berkdemir, A.; Castro-Beltran, A.; Gutiérrez, H. R.; Feng, S.; Lv, R.; Hayashi, T.; Lõpez-Urías, F.; Ghosh, S.; Muchharla, B.; Talapatra, S.; Terrones, H.; Terrones, M. Photosensor device based on few-layered WS2 films. *Adv. Funct. Mater.* **2013**, *23*, 5511–5517, doi:10.1002/adfm.201300760.

37. Koppens, F. H. L.; Mueller, T.; Avouris, P.; Ferrari, A. C.; Vitiello, M. S.; Polini, M. Photodetectors based on graphene, other two-dimensional materials and hybrid systems. *Nat. Nanotechnol.* **2014**, *9*, 780–793, doi:10.1038/nnano.2014.215.

38. Yin, Z.; Li, H. H.; Li, H. H.; Jiang, L.; Shi, Y.; Sun, Y.; Lu, G.; Zhang, Q.; Chen, X.; Zhang, H. Single-layer MoS 2 phototransistors. *ACS Nano* **2012**, *6*, 74–80, doi:10.1021/nn2024557.

39. Sun, Z.; Martinez, A.; Wang, F. Optical modulators with 2D layered materials. *Nat. Photonics* **2016**, *10*, 227–238, doi:10.1038/nphoton.2016.15.

40. Lopez-Sanchez, O.; Alarcon Llado, E.; Koman, V.; Fontcuberta I Morral, A.; Radenovic, A.;

Kis, A. Light generation and harvesting in a van der waals heterostructure. *ACS Nano* **2014**, *8*, 3042–3048, doi:10.1021/nn500480u.

41. Withers, F.; Del Pozo-Zamudio, O.; Mishchenko, A.; Rooney, A. P.; Gholinia, A.; Watanabe, K.; Taniguchi, T.; Haigh, S. J.; Geim, A. K.; Tartakovskii, A. I.; Novoselov, K. S. Light-emitting diodes by band-structure engineering in van der Waals heterostructures. *Nat. Mater.* **2015**, *14*, 301–306, doi:10.1038/nmat4205.

42. Liu, C. H.; Clark, G.; Fryett, T.; Wu, S.; Zheng, J.; Hatami, F.; Xu, X.; Majumdar, A. Nanocavity integrated van der Waals heterostructure light-emitting tunneling diode. *Nano Lett.* **2017**, *17*, 200–205, doi:10.1021/acs.nanolett.6b03801.

43. He, Y.-M.; Clark, G.; Schaibley, J. R.; He, Y.; Chen, M.-C.; Wei, Y.-J.; Ding, X.; Zhang, Q.; Yao, W.; Xu, X.; Lu, C.-Y.; Pan, J.-W. Single quantum emitters in monolayer semiconductors. *Nat. Nanotechnol.* **2015**, *10*, 497–502, doi:10.1038/nnano.2015.75.

44. Tonndorf, P.; Schmidt, R.; Schneider, R.; Kern, J.; Buscema, M.; Steele, G. A.; Castellanos-Gomez, A.; van der Zant, H. S. J.; Michaelis de Vasconcellos, S.; Bratschitsch, R. Single-photon emission from localized excitons in an atomically thin semiconductor. *Optica* **2015**, *2*, 347, doi:10.1364/OPTICA.2.000347.

45. M. Wang, A. Krasnok, T. Zhang, L. Scarabelli, L. M. Liz-Marzán, M. Terrones, A. Alù, Y. Z. Tunable Fano Resonance and Plasmon-Exciton Coupling in Single Au Nanotriangles on Monolayer WS2 at Room Temperature. *Adv. Mater.* **2018**, *in press*.

46. Wang, Q. H.; Kalantar-Zadeh, K.; Kis, A.; Coleman, J. N.; Strano, M. S. Electronics and optoelectronics of two-dimensional transition metal dichalcogenides. *Nat. Nanotechnol.* **2012**, *7*, 699–712, doi:10.1038/nnano.2012.193.

47. Amani, M.; Lien, D. H.; Kiriya, D.; Xiao, J.; Azcatl, A.; Noh, J.; Madhvapathy, S. R.; Addou, R.; Santosh, K. C.; Dubey, M.; Cho, K.; Wallace, R. M.; Lee, S. C.; He, J. H.; Ager, J. W.; Zhang, X.; Yablonovitch, E.; Javey, A. Near-unity photoluminescence quantum yield in MoS2. *Science (80-. ).* **2015**, *350*, 1065–1068, doi:10.1126/science.aad2114.

48. Amani, M.; Taheri, P.; Addou, R.; Ahn, G. H.; Kiriya, D.; Lien, D. H.; Ager, J. W.; Wallace, R. M.; Javey, A. Recombination Kinetics and Effects of Superacid Treatment in Sulfur- and Selenium-Based Transition Metal Dichalcogenides. *Nano Lett.* **2016**, *16*, 2786–2791, doi:10.1021/acs.nanolett.6b00536.

49. Johnson, A. D.; Cheng, F.; Tsai, Y.; Shih, C. K. Giant Enhancement of Defect-Bound Exciton Luminescence and Suppression of Band-Edge Luminescence in Monolayer WSe2-Ag Plasmonic Hybrid Structures. *Nano Lett.* **2017**, *17*, 4317–4322, doi:10.1021/acs.nanolett.7b01364.

50. Li, Z.; Li, Y.; Han, T.; Wang, X.; Yu, Y.; Tay, B.; Liu, Z.; Fang, Z. Tailoring MoS2 Exciton-Plasmon Interaction by Optical Spin-Orbit Coupling. *ACS Nano* **2017**, *11*, 1165–1171, doi:10.1021/acsnano.6b06834.

51. Abid, I.; Chen, W.; Yuan, J.; Bohloul, A.; Najmaei, S.; Avendano, C.; Péchou, R.; Mlayah, A.; Lou, J. Temperature-Dependent Plasmon–Exciton Interactions in Hybrid Au/MoSe 2 Nanostructures. *ACS Photonics* **2017**, *4*, 1653–1660, doi:10.1021/acsphotonics.6b00957.

52. Sigle, D. O.; Mertens, J.; Herrmann, L. O.; Bowman, R. W.; Ithurria, S.; Dubertret, B.; Shi, Y.; Yang, H. Y.; Tserkezis, C.; Aizpurua, J.; Baumberg, J. J. Monitoring morphological changes in 2D monolayer semiconductors using atom-thick plasmonic nanocavities. *ACS Nano* **2015**, *9*, 825–830, doi:10.1021/nn5064198.

53. Zhang, X.; Choi, S.; Wang, D.; Naylor, C. H.; Johnson, A. T. C.; Cubukcu, E. Unidirectional Doubly Enhanced MoS 2 Emission via Photonic Fano Resonances. *Nano Lett.* **2017**, *17*, 6715–6720, doi:10.1021/acs.nanolett.7b02777.

54. Lee, M.-G.; Yoo, S.; Kim, T.; Park, Q.-H. Large-area plasmon enhanced two-dimensional MoS 2. *Nanoscale* **2017**, *9*, 16244–16248, doi:10.1039/C7NR04974A.

55. Gonçalves, P. A. D.; Bertelsen, L. P.; Xiao, S.; Mortensen, N. A. Plasmon-exciton polaritons in two-dimensional semiconductor/metal interfaces. *Phys. Rev. B* **2018**, *97*, 041402, doi:10.1103/PhysRevB.97.041402.

56. Zheng, D.; Zhang, S.; Deng, Q.; Kang, M.; Nordlander, P.; Xu, H. Manipulating coherent plasmon-exciton interaction in a single silver nanorod on monolayer WSe2. *Nano Lett.* **2017**, *17*, 3809–3814, doi:10.1021/acs.nanolett.7b01176.

57. Huang, J.; Akselrod, G. M.; Ming, T.; Kong, J.; Mikkelsen, M. H. Tailored Emission Spectrum of 2D Semiconductors Using Plasmonic Nanocavities. *ACS Photonics* **2018**, *5*, 552–558, doi:10.1021/acsphotonics.7b01085.

58. Eda, G.; Maier, S. A. Two-dimensional crystals: Managing light for optoelectronics. *ACS Nano* **2013**, *7*, 5660–5665, doi:10.1021/nn403159y.

59. Akselrod, G. M.; Ming, T.; Argyropoulos, C.; Hoang, T. B.; Lin, Y.; Ling, X.; Smith, D. R.; Kong, J.; Mikkelsen, M. H. Leveraging nanocavity harmonics for control of optical processes in 2d semiconductors. *Nano Lett.* **2015**, *15*, 3578–3584, doi:10.1021/acs.nanolett.5b01062.

60. Tahersima, M. H.; Birowosuto, M. D.; Ma, Z.; Coley, W. C.; Valentin, M. D.; Naghibi Alvillar, S.; Lu, I.-H.; Zhou, Y.; Sarpkaya, I.; Martinez, A.; Liao, I.; Davis, B. N.; Martinez, J.; Martinez-Ta, D.; Guan, A.; Nguyen, A. E.; Liu, K.; Soci, C.; Reed, E.; Bartels, L.; Sorger, V. J. Testbeds for Transition Metal Dichalcogenide Photonics: Efficacy of Light Emission Enhancement in Monomer vs Dimer Nanoscale Antennae. *ACS Photonics* **2017**, *4*, 1713–1721, doi:10.1021/acsphotonics.7b00208.

61. Noori, Y. J.; Cao, Y.; Roberts, J.; Woodhead, C.; Bernardo-Gavito, R.; Tovee, P.; Young, R. J. Photonic Crystals for Enhanced Light Extraction from 2D Materials. *ACS Photonics* **2016**, *3*, 2515–2520, doi:10.1021/acsphotonics.6b00779.

62. Galfsky, T.; Sun, Z.; Considine, C. R. C. R.; Chou, C.-T. T.; Ko, W.-C. C.; Lee, Y.-H. H.; Narimanov, E. E.; Menon, V. M. Broadband Enhancement of Spontaneous Emission in Two-Dimensional Semiconductors Using Photonic Hypercrystals. *Nano Lett.* **2016**, *16*, 4940–4945, doi:10.1021/acs.nanolett.6b01558.

63. Wen, J.; Wang, H.; Wang, W.; Deng, Z.; Zhuang, C.; Zhang, Y.; Liu, F.; She, J.; Chen, J.; Chen, H.; Deng, S.; Xu, N. Room-Temperature Strong Light–Matter Interaction with Active Control in Single Plasmonic Nanorod Coupled with Two-Dimensional Atomic Crystals. *Nano Lett.* **2017**, *17*, 4689–4697, doi:10.1021/acs.nanolett.7b01344.

64. Wang, S.; Li, S.; Chervy, T.; Shalabney, A.; Azzini, S.; Orgiu, E.; Hutchison, J. A.; Genet, C.; Samorì, P.; Ebbesen, T. W. Coherent coupling of WS2 monolayers with metallic photonic nanostructures at room temperature. *Nano Lett.* **2016**, *16*, 4368–4374, doi:10.1021/acs.nanolett.6b01475.

65. Baranov, D. G.; Wersäll, M.; Cuadra, J.; Antosiewicz, T. J.; Shegai, T. Novel Nanostructures and Materials for Strong Light–Matter Interactions. *ACS Photonics* **2018**, *5*, 24–42, doi:10.1021/acsphotonics.7b00674.

66. Wang, M.; Li, W.; Scarabelli, L.; Rajeeva, B. B.; Terrones, M.; Liz-Marzán, L. M.; Akinwande, D.; Zheng, Y. Plasmon–trion and plasmon–exciton resonance energy transfer from a single plasmonic nanoparticle to monolayer MoS2. *Nanoscale* **2017**, *9*, 13947–13955, doi:10.1039/C7NR03909C.

67. Jones, A. M.; Yu, H.; Ghimire, N. J.; Wu, S.; Aivazian, G.; Ross, J. S.; Zhao, B.; Yan, J.; Mandrus, D. G.; Xiao, D.; Yao, W.; Xu, X. Optical generation of excitonic valley coherence in monolayer WSe 2. *Nat. Nanotechnol.* **2013**, *8*, 634–638, doi:10.1038/nnano.2013.151.

68. Mak, K. F.; McGill, K. L.; Park, J.; McEuen, P. L. The valley Hall effect in MoS2 transistors. *Science (80-. ).* **2014**, *344*, 1489–1492, doi:10.1126/science.1250140.

69. Lagarde, D.; Bouet, L.; Marie, X.; Zhu, C. R.; Liu, B. L.; Amand, T.; Tan, P. H.; Urbaszek, B. Carrier and Polarization Dynamics in Monolayer MoS2. *Phys. Rev. Lett.* **2014**, *112*, 047401, doi:10.1103/PhysRevLett.112.047401.

70. Mak, K. F.; He, K.; Shan, J.; Heinz, T. F. Control of valley polarization in monolayer MoS2 by optical helicity. *Nat. Nanotechnol.* **2012**, *7*, 494–498, doi:10.1038/nnano.2012.96.

71. Onga, M.; Zhang, Y.; Ideue, T.; Iwasa, Y. Exciton Hall effect in monolayer MoS2. *Nat. Mater.* **2017**, *16*, 1193–1197, doi:10.1038/nmat4996.

72. Shin, D.; Hübener, H.; De Giovannini, U.; Jin, H.; Rubio, A.; Park, N. Phonon-driven spin-Floquet magneto-valleytronics in MoS2. *Nat. Commun.* **2018**, *9*, 638, doi:10.1038/s41467-018-02918-5.

73. Zhu, C. R.; Zhang, K.; Glazov, M.; Urbaszek, B.; Amand, T.; Ji, Z. W.; Liu, B. L.; Marie, X. Exciton valley dynamics probed by Kerr rotation in WSe2 monolayers. *Phys. Rev. B - Condens. Matter Mater. Phys.* **2014**, *90*, 1–5, doi:10.1103/PhysRevB.90.161302.

74. Plechinger, G.; Nagler, P.; Arora, A.; Schmidt, R.; Chernikov, A.; del Águila, A. G.; Christianen, P. C. M.; Bratschitsch, R.; Schüller, C.; Korn, T. Trion fine structure and coupled

spin–valley dynamics in monolayer tungsten disulfide. *Nat. Commun.* **2016**, *7*, 12715, doi:10.1038/ncomms12715.

75. Molina-Sánchez, A.; Sangalli, D.; Wirtz, L.; Marini, A. Ab Initio Calculations of Ultrashort Carrier Dynamics in Two-Dimensional Materials: Valley Depolarization in Single-Layer WSe$_2$. *Nano Lett.* **2017**, *17*, 4549–4555, doi:10.1021/acs.nanolett.7b00175.

76. Selig, M.; Berghäuser, G.; Raja, A.; Nagler, P.; Schüller, C.; Heinz, T. F.; Korn, T.; Chernikov, A.; Malic, E.; Knorr, A. Excitonic linewidth and coherence lifetime in monolayer transition metal dichalcogenides. *Nat. Commun.* **2016**, *7*, doi:10.1038/ncomms13279.

77. Aharonovich, I.; Englund, D.; Toth, M. Solid-state single-photon emitters. *Nat. Photonics* **2016**, *10*, 631–641, doi:10.1038/nphoton.2016.186.

78. Steinleitner, P.; Merkl, P.; Nagler, P.; Mornhinweg, J.; Schüller, C.; Korn, T.; Chernikov, A.; Huber, R. Direct Observation of Ultrafast Exciton Formation in a Monolayer of WSe2. *Nano Lett.* **2017**, *17*, 1455–1460, doi:10.1021/acs.nanolett.6b04422.

79. Zhu, C. R.; Zhang, K.; Glazov, M.; Urbaszek, B.; Amand, T.; Ji, Z. W.; Liu, B. L.; Marie, X. Exciton valley dynamics probed by Kerr rotation in WSe2 monolayers. *Phys. Rev. B* **2014**, *90*, 161302, doi:10.1103/PhysRevB.90.161302.

80. Glazov, M. M.; Amand, T.; Marie, X.; Lagarde, D.; Bouet, L.; Urbaszek, B. Exciton fine structure and spin decoherence in monolayers of transition metal dichalcogenides. *Phys. Rev. B* **2014**, *89*, 201302, doi:10.1103/PhysRevB.89.201302.

81. Molina-Sánchez, A.; Sangalli, D.; Wirtz, L.; Marini, A. Ab Initio Calculations of Ultrashort Carrier Dynamics in Two-Dimensional Materials: Valley Depolarization in Single-Layer WSe$_2$. *Nano Lett.* **2017**, *17*, 4549–4555, doi:10.1021/acs.nanolett.7b00175.

82. Petersen, J.; Volz, J.; Rauschenbeutel, A. Chiral nanophotonic waveguide interface based on spin-orbit interaction of light. *Science (80-. ).* **2014**, *346*, 67–71, doi:10.1126/science.1257671.

83. Rodriguez-Fortuno, F. J.; Marino, G.; Ginzburg, P.; O'Connor, D.; Martinez, A.; Wurtz, G. A.; Zayats, A. V. Near-Field Interference for the Unidirectional Excitation of Electromagnetic Guided Modes. *Science (80-. ).* **2013**, *340*, 328–330, doi:10.1126/science.1233739.

84. Li, S. V.; Baranov, D. G.; Krasnok, A. E.; Belov, P. A. All-dielectric nanoantennas for unidirectional excitation of electromagnetic guided modes. *Appl. Phys. Lett.* **2015**, *107*, doi:10.1063/1.4934757.

85. Lodahl, P.; Mahmoodian, S.; Stobbe, S.; Schneeweiss, P.; Volz, J.; Rauschenbeutel, A.; Pichler, H.; Zoller, P. Chiral Quantum Optics. *Nature* **2016**, *541*, 473–480, doi:10.1038/nature21037.

86. Baranov, D. G.; Wersäll, M.; Cuadra, J.; Antosiewicz, T. J.; Shegai, T. Novel Nanostructures and Materials for Strong Light–Matter Interactions. *ACS Photonics* **2017**, acsphotonics.7b00674, doi:10.1021/acsphotonics.7b00674.

87. Krasnok, A.; Caldarola, M.; Bonod, N.; Alú, A. Spectroscopy and Biosensing with Optically Resonant Dielectric Nanostructures. *Adv. Opt. Mater.* **2018**, *6*, 1701094, doi:10.1002/adom.201701094.